\documentclass[aps,longbibliography,prd,twocolumn,groupedaddress]{revtex4}
\newcommand{\be}{\begin{equation}}
\newcommand{\ee}{\end{equation}}
\usepackage{graphicx}
\usepackage{amsmath,amssymb,amsfonts}
\usepackage[colorlinks,citecolor=blue,linkcolor=blue,urlcolor=blue]{hyperref}
\usepackage[applemac]{inputenc}
\newcommand{\dd}{\mathrm{d}}
\newcommand{\sign}[1]{\text{sign\,} #1}
\newcommand{\artanh}[1]{\text{arctanh} #1}
\newcommand{\arsinh}[1]{\text{arnsinh} #1}
\newcommand{\PD}[2]{\ensuremath{\frac{\partial #1}{\partial #2}}}

\begin{document}

\title{How Information Crosses Schwarzschild's Central Singularity}

\author{Fabio D'Ambrosio}\email{fabio.dambrosio@gmx.ch}
\author{Carlo Rovelli}\email{rovelli@cpt.univ-mrs.fr}

\affiliation{\small
\mbox{CPT, Aix--Marseille Universit\'e, Universit\'e de Toulon, CNRS, Case 907, F--13288 Marseille, France.} }
\date{\small\today}
\begin{abstract}
\noindent
We study the natural extension of spacetime across Schwarzschild's central  singularity and the behavior of the geodesics crossing it. Locality implies that this extension is independent from the future fate of black holes.  We argue that this extension is the natural $\hbar\!\to\!0$ limit of the effective quantum geometry inside a black hole, and show that the central region contains causal diamonds with area satisfying Bousso's bound for an entropy that can  be as large as Hawking's radiation entropy. This result sheds light on the possibility that Hawking radiation is purified by information crossing the internal singularity. 
\end{abstract}

\maketitle 

\section{Non-Riemannian Extension}\label{sec:NonRiemannianExtension} 

Einstein cautioned repeatedly against giving excessive weight to the fact that the gravitational field determines a (pseudo-) Riemannian geometry~\cite{Lehmkuhl2014a}.  He regarded this fact as a convenient mathematical feature and a tool to connect the theory to the geometry of Newton's and Minkowski's spaces~\cite{Einstein_1921}, but the essential point about~$g_{\mu\nu}$ is not that it describes gravitation as a manifestation of a Riemannian geometry; it is that it provides a relativistic field theoretical description of gravitation~\cite{Einstein_Meaning}.  Well behaved solutions of the field equations might thus be physically relevant even when they fail to define a geometry which is --strictly speaking-- a Riemannian manifold. 

This consideration is relevant for understanding the interior of black holes.  There is no Riemannian manifold extending the Schwarzschild metric beyond the central singularity where the Schwarzschild radius vanishes:~$r_s=0$. There is indeed abundant mathematical literature about the inextensibility {\em in this sense} and the related geodesic incompleteness of the Schwarzschild spacetime (see~\cite{Hawking_Ellis,Kriele,Sbierski:2015} for instance). \emph{But} there is a smooth solution of the equations that continues across~$r_s=0$.  It defines a metric geometry that is Riemannian almost everywhere, with curvature invariants diverging on a low dimensional surface.  The metric geometry defined by this extension continues the interior of the black hole across~$r_s=0$ into the geometry of the interior of a white hole.  

This possibility was noticed by several authors over the past decades.  To the best of our knowledge it was first reported by Synge in the fifties~\cite{Synge1950} and rediscovered by Peeters, Schweigert and van~Holten in the nineties~\cite{Peeters:1994jz}. A similar observation has recently been made in the context of cosmology in~\cite{Koslowski:2016hds}.  Here we study this extension and all geodesics that cross~$r_s=0$.

This geometry can be seen as the~$\hbar\to0$ limit of an effective metric determined by quantum gravity. On physical grounds we expect what happens near~$r_s\!=\!0$ to be affected by quantum effects, because curvature reaches the Planck scale in this region.    

Notice that quantum gravity is expected to render what happens at distances smaller than the Planck length physically irrelevant~\cite{Rovelli1994a}, therefore curvature singularities on low dimensional surfaces are likely to be physically meaningless anyway.  The possibility of a quantum transitions across~$r_s=0$ has been indeed explored by many authors, see for instance~\cite{Modesto2006,Modesto2008,Hossenfelder:2009fc}.  

Quantum gravity is also expected to bound curvature~\cite{Narlikar1974,Frolov:1979tu,Frolov:1981mz,Stephens1994,modesto2004disappearance,Mazur:2004, Ashtekar:2005cj, Balasubramanian:2006,Hayward2006, Hossenfelder:2009fc,Hossenfelder:2010a,frolov:BHclosed, Rovelli2013d, Bardeen2014,Giddings1992a,Giddings1992b,Hossenfelder:2012uy}.   If we assume that the curvature of the effective metric is bound at the Planck scale, the central singularity is crossed by a regular (pseudo-) Riemmannian metric \emph{without} singular regions. Below we write an explicit ansatz for such an effective metric. 

The quantum bound on the curvature determines the size $l$ of its minimal surface (the ``Planck Star", where the geometry bounces) to be of order  $l\sim m^{\frac13}$ in Planck units~\cite{Rovelli2014}. We show that the central region of a black hole contains causal diamonds with equators having large area.  In the case of a black hole of initial mass $m$ evaporating in a time $\sim m^3$, this area can be as large as 
\be
       	A \sim 2\pi \sqrt{2ml}\ m^3 \gg 16\pi m^2.
\ee
According to Bousso's covariant bound  \cite{Bousso1999}, this region of spacetime is sufficiently large to contain an entropy of the same order as the entropy of Hawking radiation.  

This result supports the idea that Hawking radiation is purified by information that crosses the central singularity when a black hole quantum tunnels into a white hole \cite{noi}. 

\section{The $A$ Region inside a Black Hole}\label{sec:TheARegion} 

\begin{figure}[h]
	\centering
\includegraphics[height=4cm]{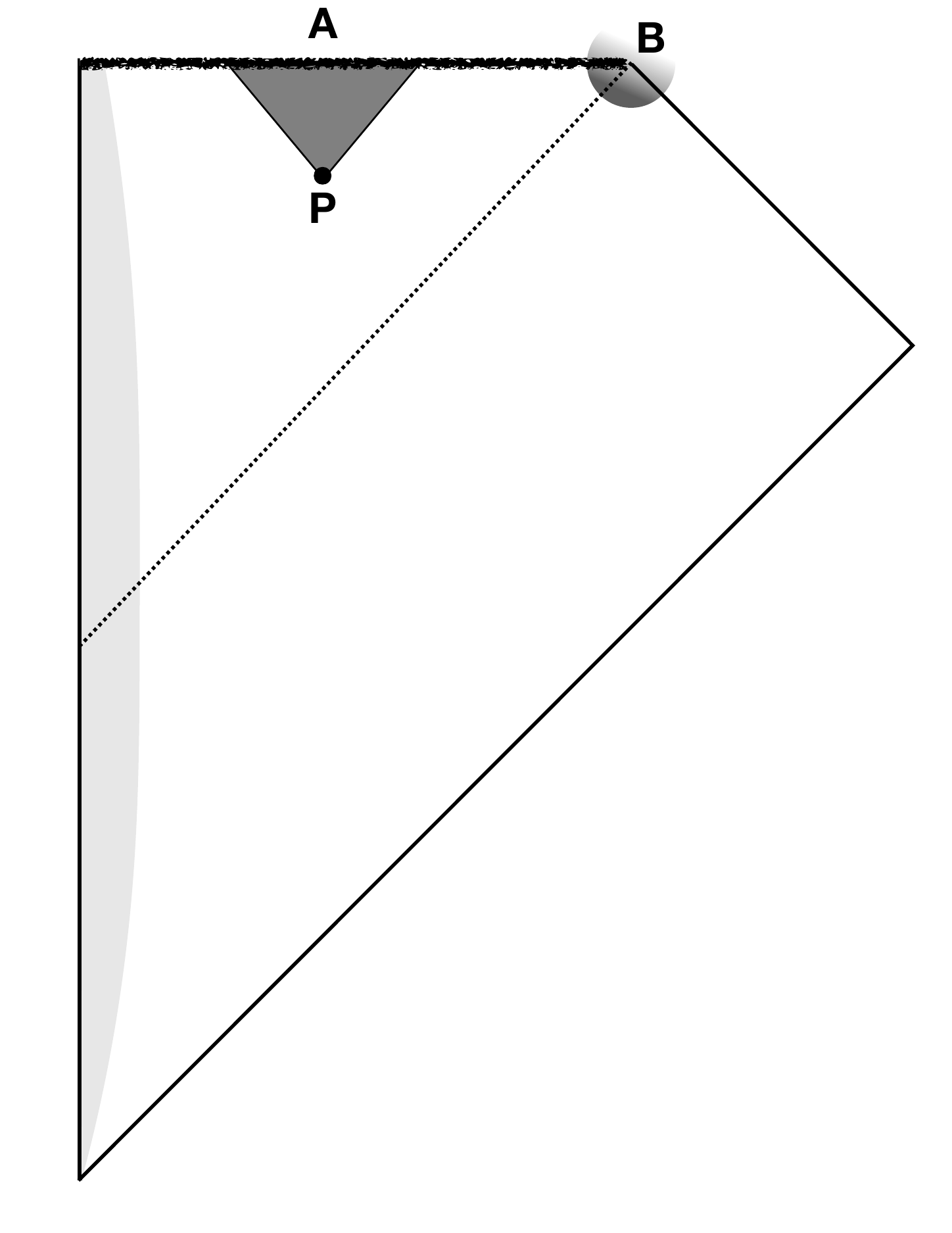}
\caption{\em The conformal diagram of the spacetime of a collapsing star predicted by classical GR. The star is light grey, the horizon is dotted, the $r_s=0$ singularity is the upper thick line.}
\label{uno}
\end{figure}

Figure~\ref{uno} represents the standard Carter-Penrose conformal diagram of a star that collapses in classical General Relativity, disregarding any quantum effects.  

We pick a generic point~$P$ inside the hole and we are interested in its future, in particular what happens past the upper line of the figure, which is the central Schwarzschild~$r_s=0$ singularity.    It is important to notice that this region is causally disconnected from the region indicated as~$B$ in the conformal diagram, which is the region relevant for the long term future of the black hole. Region~$B$ is going to be substantially affected by Hawking evaporation, possible final disappearance of the black hole, and the like. We are studying all this elsewhere~\cite{noi}. But nothing of this concerns what happens in the future of~$P$ near the singularity, {\em because this is causally disconnected from~$B$}.  

We call the local transition that we study here, unaffected by the long term behavior of the hole, ``region~$A$". 

\begin{figure}
\includegraphics[height=3cm]{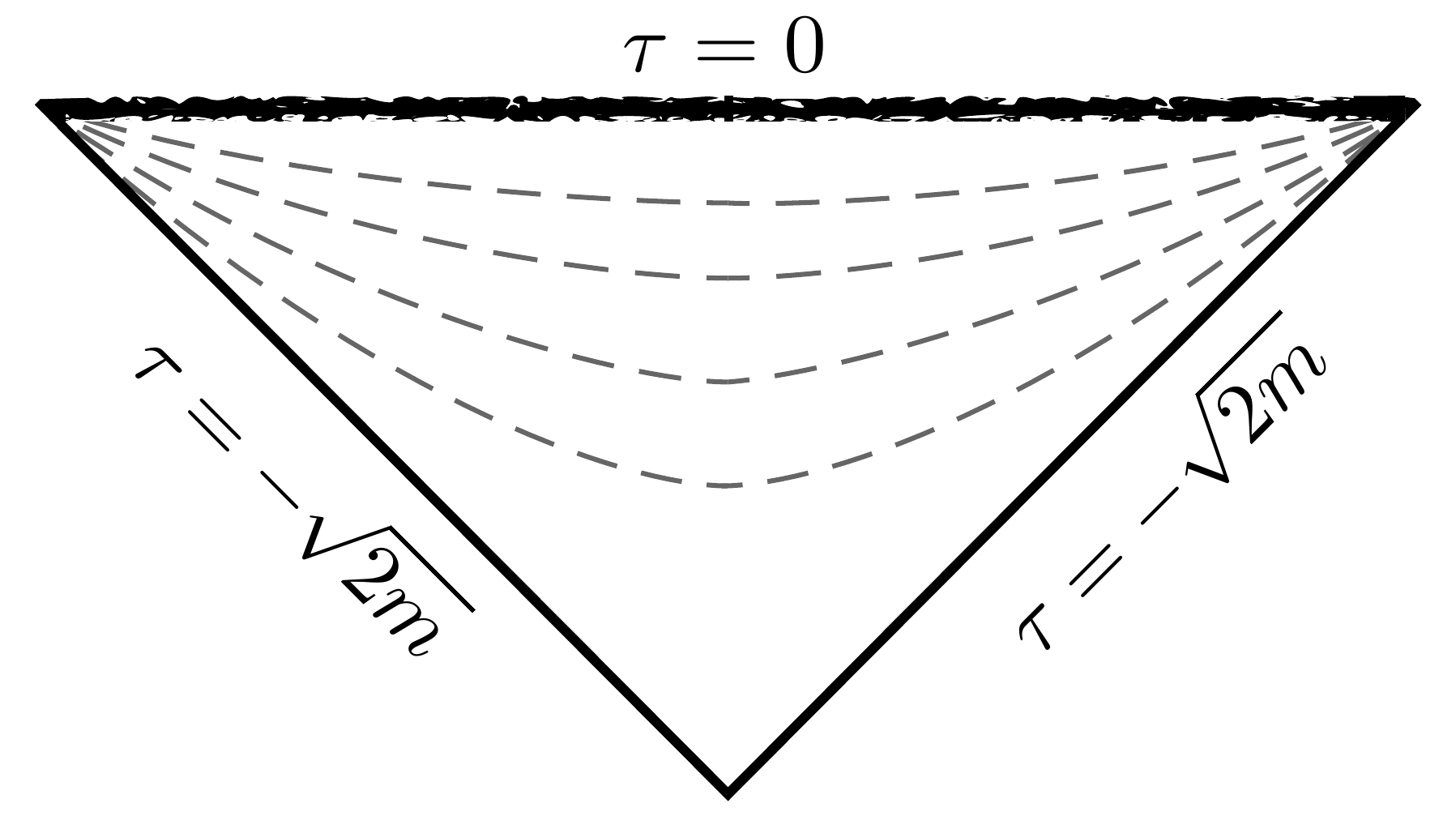}
\vspace{-1em}
\caption{\em Interior of black hole with (space-like) constant $\tau$ (or constant Schwarzschild radius) surfaces.}
\label{lowerBH}
\end{figure}

To study this region, let us write the metric explicitly.  The interior of a Schwarzschild black hole is spherically symmetric and homogeneous in a third spacial direction, which we coordinatize with a space-like coordinate~$x$.  (Which is the Schwarzschild coordinate~$t_s$ that becomes space-like inside the horizon.)  Therefore, it can be foliated by space-like surfaces that have each the geometry of a 3d cylinder. A sphere times the real line.   By spherical symmetry, and homogeneity along the~$x$~coordinate, the  gravitational field~$g_{\mu\nu}(\tau,x,\theta,\phi)$ can be written in the form
\be
\dd s^2=g_{\tau\tau}(\tau)\dd\tau^2-g_{xx}(\tau)\dd x^2-g_{\theta\theta}(\tau)\dd\Omega^2,
\ee
where $\dd\Omega^2=\dd\theta^2+\sin^2\!\theta\, \dd\phi^2$ is the metric of the unit sphere. The coordinates $\theta\in[0,\pi]$ and $\phi\in[0,2\pi[$ are standard  coordinates on the sphere. The coordinate $x\in\ ]x_{\textsf{min}},x_{\textsf{max}}[$ runs along an arbitrary finite portion of the cylinder's axis, and $\tau$ is a temporal coordinate, whose range we will explore in studying the dynamics.  Inserting this field in the Einstein equations we find the solution  
\be
g_{\tau\tau}(\tau)=\frac{4\tau^4}{2m-\tau^2},\ \ 
g_{xx}(\tau)=\frac{2m-\tau^2}{\tau^2},\ \ 
g_{\theta\theta}(\tau)=\tau^4.
\nonumber \label{solution}
\ee
The value $\tau\!=\!0$ locates where the cylinder's radius shrinks to zero. The corresponding line element is
\be
\dd s^2=\frac{4\tau^4}{2m-\tau^2}\dd\tau^2-\frac{2m-\tau^2}{\tau^2}\dd x^2-\tau^4\dd\Omega^2. \label{met}
\ee
The region $-\sqrt{2m}<\tau<0$ is precisely the standard interior of a black hole, namely region II of the Kruskal extension of the Schwarzschild solution. This can be seen by going to the usual Schwarzschild coordinates
\be
t_s=x \ \quad \text{and} \quad \ r_s=\tau^2,
\ee
which puts the metric in the usual Schwarzschild form
\be
\dd s^2=\left(1-\frac{2m}{r_s}\right)\dd t_s^2-\left(1-\frac{2m}{r_s}\right)^{-1}\!\!\dd r_s^2-r_s^2\dd\Omega^2.
\ee 
This line element, as is well known, solves the Einstein equations also in the region $r_s<2m$ where it describes the black hole interior.  As $\tau$ flows from $-\sqrt{2m}$ to zero, the Schwarzschild radius shrinks from the horizon to the central singularity. The resulting geometry is depicted in Figure~\ref{lowerBH}, for the full range $x\in\ ]-\infty,+\infty[$. The divergence at $\tau=0$ is the central black hole singularity at $r_s=0$.

But notice the following. Differential equations can develop fake singularities because they are formulated in inconvenient variables. For instance, a solution of the equation $ 
y\ddot y-2\dot y^2+y^2=0$, is $y(t)={1}/{\sin t}$ which diverges at $t=0$. However, by simply defining $x={1}/{y}$, the differential equation turns into the familiar $\ddot x=-x$ whose solution $x=\sin t$ is regular across $t=0$. 

The same can be done for the back hole interior. Let us change variables from the three variables $g_{\tau\tau}, g_{xx},$ and  $g_{\theta\theta}$ to the three variables $a, b,$ and $N$ defined by~\cite{Kenmoku1997}
\be
 g_{\tau\tau}= N^2\; \frac{a}{b},  \ \ \ g_{xx} =\frac{b}{a}, \quad \text{and} \quad \ g_{\theta\theta}=a^2.
\ee
This is a change of dynamical (configuration) variables, not to be confused with a coordinate transformation, namely with a change of the independent parameters $(\tau,x,\theta,\phi)$. 
Inserting these new variables into the first order action of General Relativity yields
\be
S=\frac{v}{4G}\int \dd\tau \left(N-\frac{\dot a\dot b}{N}\right),\label{action}
\ee
where $v=\int_{x_{\textsf{min}}}^{x_{\textsf{max}}} \dd x$ and $G$ is Newton's constant. The equations of motion of this action are 
\be
\frac{\dd}{\dd\tau}\frac{\dot a}{N}=0, \ \quad\ \frac{\dd}{\dd\tau}\frac{\dot b}{N}=0, \quad\  \text{and} \quad \ \dot a  \dot b+N^2=0.
\ee
They are solved in particular by
\be 
a(\tau)=\tau^2, \ \quad \ b(\tau)=2m-\tau^2, \ \quad\   N^2(\tau)=4\,a(\tau).\label{sol}
\ee
This gives precisely the solution~\eqref{solution}, namely the black hole interior. So far, we have only done a consistent change of variables in a dynamical system. 

But now it is evident from equation~\eqref{sol} that the solution can be continued past $\tau=0$ without any loss of regularity.  Expressed in terms of these variables, the gravitational field evolves regularly past the central singularity of a black hole, to positive values of $\tau$.

\begin{figure}[h]
\includegraphics[height=4.2cm]{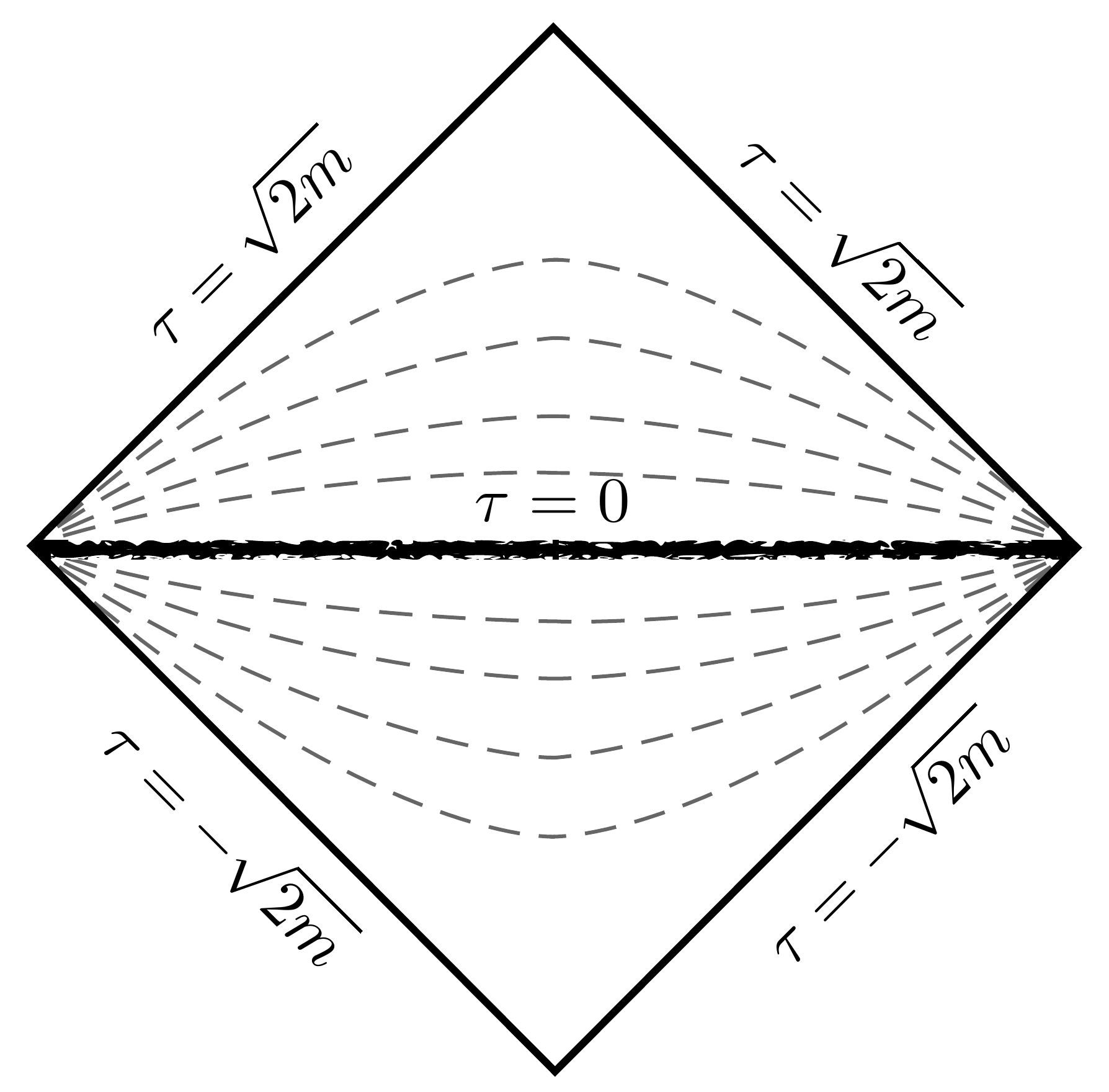}
\caption{\small \em The interior transition across the $A$ region.}
\label{4}
\end{figure}

For positive values of $\tau$ the geometry determined by this solution of the gravitational field equations is simply the time reversal of the black hole interior, namely a white hole interior, joined to the black hole across the singularity, as depicted in Figure~\ref{4}. 

The geometry defined in this way is given by the line element~\eqref{met} where the coordinate $\tau$ covers the full range $-\sqrt{2m}<\tau<\sqrt{2m}$. 

For positive and for negative $\tau$ this line element defines a Ricci flat pseudo-Riemannian geometry.  Not so for $\tau=0$ where --for instance-- the scalar $K^2\sim R_{\mu\nu\rho\sigma}R^{\mu\nu\rho\sigma}$ constructed by squaring the Riemann tensor, diverges as
\be
K(\tau)\sim \frac{m}{\tau^6}. \label{dive}
\ee
Because of this divergence, this spacetime is not a Riemannian manifold. However, it is still a metric manifold and it can be approximated with arbitrary precision by a genuine (pseudo-) Riemannian manifold.  

More precisely, we can can view the metric~\eqref{met} as a ``distributional Riemannian geometry", in the following sense. We say that a distributional Riemannian geometry $\dd s$ on a manifold is the assignment of a length $L[\gamma]$ to any curve on the manifold, such that there is a one-parameter family of Riemannian geometries $\dd s_l$ such that $\lim_{l\to 0}\int_\gamma \dd s_l=L[\gamma]$.  The metric~\eqref{met} is a distributional geometry in this sense.   

In the Section \ref{cd} we give an explicit example of  a one parameter family of Riemannian metrics $\dd s_l$ converging to the metric~\eqref{met} and we argue that $\dd s_l$ can have a direct physical interpretation in quantum gravity.   Before this, in the next section we study the geodesics that cross the singularity for the line element~\eqref{met}.

\section{Geodesics crossing $r_s=0$}\label{sec:GeodesicsCrossing}

We study the geodesics of the metric described above using the relativistic Hamilton-Jacobi formalism. An advantage of this method is that it does not require us to think in terms of  evolution of the coordinates as functions of an unphysical parameter; rather, it gives us directly the physical worldline in terms of coordinates as functions of one another.   It gives us directly a gauge invariant expression for the geodesic. 

The relativistic Hamilton-Jacobi approach requires us to find a three-parameter family of solutions to the Hamilton-Jacobi equation 
\begin{align}\label{eq:HamiltonJacobiEq}
	g^{\mu\nu}\,\PD{S}{x^\mu} \PD{S}{x^{\nu}} = \varepsilon,
\end{align}
where $S(x^{\mu},P_a)$ is Hamilton's principal function. The three parameters $P_a$, $a=1,2,3$, are integration constants and $\varepsilon = 1$ for massive particles (time-like geodesics) while $\varepsilon = 0$ for massless particles (null geodesics).   The geodesics are directly found by imposing 
\begin{align}
	\PD{S(x^\mu, P_a)}{P_a}-Q^a = 0, \label{Q}
\end{align}
where $Q^a$ are the other three integration constants.  

Due to the spherical symmetry of the Schwarzschild spacetime, angular momentum is conserved and the motions are planar. Without loss of generality we can choose spherical coordinates such that the motions lie in the equatorial plane $\theta = \frac{\pi}{2}$. This effectively reduces the problem to two dimensions. In the $\theta = \frac{\pi}{2}$ plane, the metric becomes 
\be
\dd s^2=\frac{4\tau^4}{2m-\tau^2}\dd\tau^2-\frac{2m-\tau^2}{\tau^2}\dd x^2-\tau^4 \dd\phi^2, \label{met3}
\ee 
and the Hamilton-Jacobi equation reads 
\be
\frac{2m-\tau^2}4\left(\PD{S}{\tau}\right)^2-\frac{\tau^6}{2m-\tau^2}\left(\PD{S}{x}\right)^2-\left(\PD{S}{\phi}\right)^2={\tau^4} \varepsilon. \nonumber
\ee 
Due to spherical symmetry we only need a two-parameter family of solutions. This is easy to write:
\begin{align}
	S =& P x + L \phi - 2\int \sqrt{\varepsilon\tau^4 + L^2 + \frac{P^2 \tau^6}{2m-\tau^2}}\frac{\dd \tau}{\sqrt{2m-\tau^2}}.\nonumber
\end{align}
It is parametrized by angular momentum $L$ and the conserved charge $P$ conjugate to the cyclic variable $x$. Using~\eqref{Q} we have then the following expressions for the geodesics
\begin{align}\label{eq:Motions}
	x(\tau)&=x_0 + \int\!\!\!\frac{2 P \tau^6}{(2m-\tau^2)^{\frac{3}{2}}\sqrt{\varepsilon\tau^4 + L^2 + \frac{P^2 \tau^6}{2m-\tau^2}}} \dd \tau, \notag\\
	\phi(\tau)&=\phi_0 + \int\!\!\! \frac{2L}{\sqrt{2m-\tau^2}\sqrt{\varepsilon\tau^4 + L^2 + \frac{P^2 \tau^6}{2m-\tau^2}}}\dd \tau.
\end{align}
These give the geodesic motions.  Notice that the equations of motion are well defined in $\tau=0$ since the integrands are finite. In what follows we will first uncover the physical meaning of the conserved charge $P$ and then solve the  integrals explicitly for time-like and null geodesics under different assumptions on the conserved charges $P$ and $L$.

\subsection{The physical Meaning of $S(x^\mu, P_a)$, $P$ and $L$}
Hamilton's principal function for a particle on a fixed background has a transparent physical meaning: It is equal to the particle's proper time along a given trajectory. To see this in full generality, we consider the particle's Lagrangian
\begin{align}
	L\left(q^\mu, \dot q^\mu\right) = \sqrt{g_{\mu\nu}(q) \dot q^\mu \dot q^\nu}
\end{align}
in configuration space variables $q^\mu$, $\mu=1,\dots n$. Trajectories $q^\mu = q^\mu(\lambda)$ are assumed to be arbitrarily parametrized by $\lambda$ and the dot indicates a derivative with respect to $\lambda$. As is well known, a Legendre transformation which trades the $n$ velocities $\dot q^\mu$ for the $n$ momenta $p_\mu$ leaves us with the vanishing Hamiltonian
\begin{align}
	H(q^\mu, p_\mu) = p_\mu \dot q^\mu - L(q^\mu, p_\mu) = 0.
\end{align}
A consequent canonical transformation then leads to the Hamilton-Jacobi equation
\begin{align}
	H\left(q^\mu, \frac{\partial S}{\partial q^\mu}\right) = 0,
\end{align}
which is solved by $S=S(q^\mu, P_a)$ with $\frac{\partial S}{\partial q^{a}}=P_a = const.$ for $a=1,\dots, k< n$. The particle's phase space is now coordinatized by the $n$ generalized coordinates~$q^\mu$, the $k$ constants $P_a$ and the $n-k$ momenta $\frac{\partial S}{\partial q^{i}}$ with $k<i\leq n$. For simplicity we denote the momenta collectively as $p_\mu := (P_a,  \frac{\partial S}{\partial q^{i}})$. It then follows that
\begin{align}
	\dd S(q^\mu, P_a) &= \frac{\partial S}{\partial q^\mu}\dd q^\mu = p_\mu \dd q^\mu =  P_a \dd q^{a} + \frac{\partial S}{\partial q^{i}}\dd q^{i},
\end{align}
which can be integrated along a geodesic with start and end point $q_0^\mu$ and $q^\mu$, respectively, to yield
\begin{align}
	S(q^\mu, P_a) = \int_{q_0^\mu}^{q^\mu}p_\mu \dd \tilde q^\mu =  P_a \left(q^{a}-q^{a}_0\right) + \int_{q_0^{i}}^{q^{i}}\frac{\partial S}{\partial \tilde{q}^{i}}\dd \tilde{q}^{i}.
\end{align}

This general expression is of the same form as the explicit solution found in the previous section. But notice that since the vanishing Hamiltonian implies $p_\mu \dot q^\mu = L(q^\mu, p_\mu)$, the one-form $\dd S$ can equivalently be written as
\begin{align}
	\dd S(q^\mu, P_a) = p_\mu \dd q^\mu = p_\mu \dot q^\mu \dd \lambda = L(q^\mu, p_\mu)\dd \lambda.
\end{align}
Integrating this one-form along the same geodesic as before yields
\begin{align}\label{eq:equivalence}
	S(q^\mu, P_a) &= \int_{q_0^\mu}^{q^\mu}p_\mu \dd \tilde q^\mu = \int_{\lambda_0}^{\lambda} L(q^\mu, \dot q^\mu)\dd \tilde\lambda \notag\\
	&= \int_{\lambda_0}^{\lambda}\sqrt{g_{\mu\nu}(q)\dot q^\mu\dot q^\nu}\dd \tilde\lambda.
\end{align}
That is: Hamilton's principal function is equal to the particle's proper time along a given geodesic.

This equivalence simplifies the interpretation of the conserved charges $P$ and $L$. On the right hand side of \eqref{eq:equivalence} we have the standard action for a particle on a fixed background $g_{\mu\nu}$. This action is invariant under variations of the Schwarzschild coordinates $t_s$ and $\phi$ in the $r>2m$ region, which gives rise to two conserved charges. More precisely, there are two Killing vector fields, $V=\partial_{t_s}$ and $W = \partial_\phi$, and the conserved charges can be written as
\begin{align}
	E = g_{{t_s} {t_s}} V^{t_s} \dot t_s\quad\text{and}\quad L = g_{\phi\phi} W^\phi \dot \phi.
\end{align}
To call $L$ angular momentum requires no further justification while $E$ is found to coincide with the special relativistic notion of energy when $r_s\rightarrow \infty$.

As the conserved charges are given in a manifestly coordinate independent form and we know of many gauges which extend smoothly across the horizon we reach the following conclusion. Particle trajectories in the outside region are labelled by $E$ and $L$ and a particle crossing the horizon from the outside continues on one of the inside geodesics discussed in this article, which are labelled by $P$ and $L$.   We can thus identify $P$ with the energy $E$.  

\begin{figure}[h]
	\centering
\includegraphics[height=5cm]{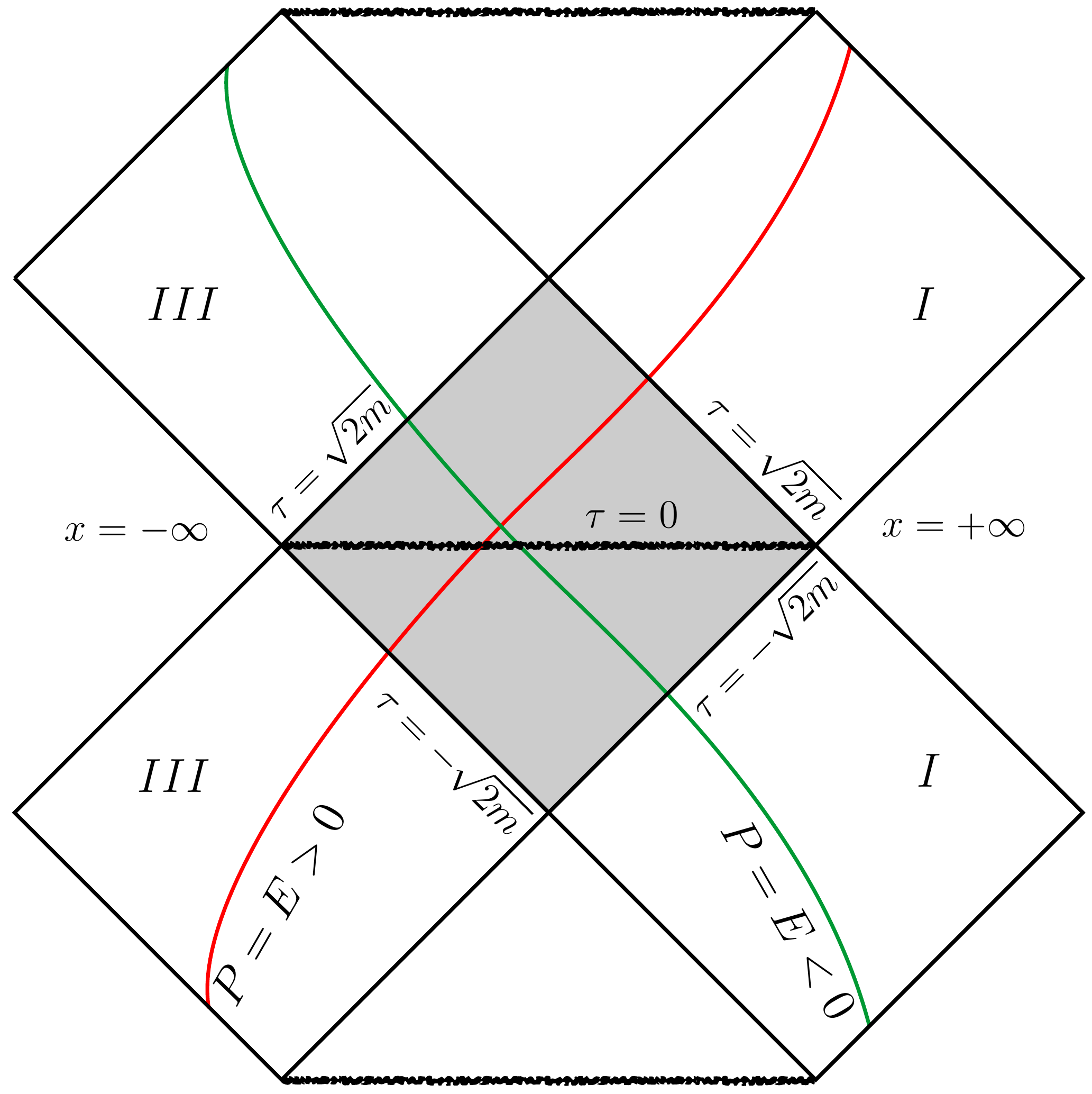}
\caption{\em Time-like geodesics with $E>0$ originate from the lower region $III$ and extend into the upper region $I$. Geodesics with $E<0$ move from the lower right to the top left.}
\label{fig:ExtSpacetime}
\end{figure}

The sign of $P$ determines whether the geodesic is moving towards decreasing or increasing $x$.  If we join the horizons $\tau=\pm\sqrt{2m}$ to two complete Kruskal spacetimes (see Figure~\ref{fig:ExtSpacetime}), time-like geodesics incoming from the lower region $III$ and emerging in the upper region $I$ have positive $P$, and $P$ can be identified with the conventional energy $E$ at $r_s\rightarrow \infty$ in this region.   $E$ is negative for the time-like geodesics moving in the opposite direction. 

\subsection{$L=0$}\label{ssc:LZero}

\begin{figure}[h]
	\centering
\includegraphics[height=4.2cm]{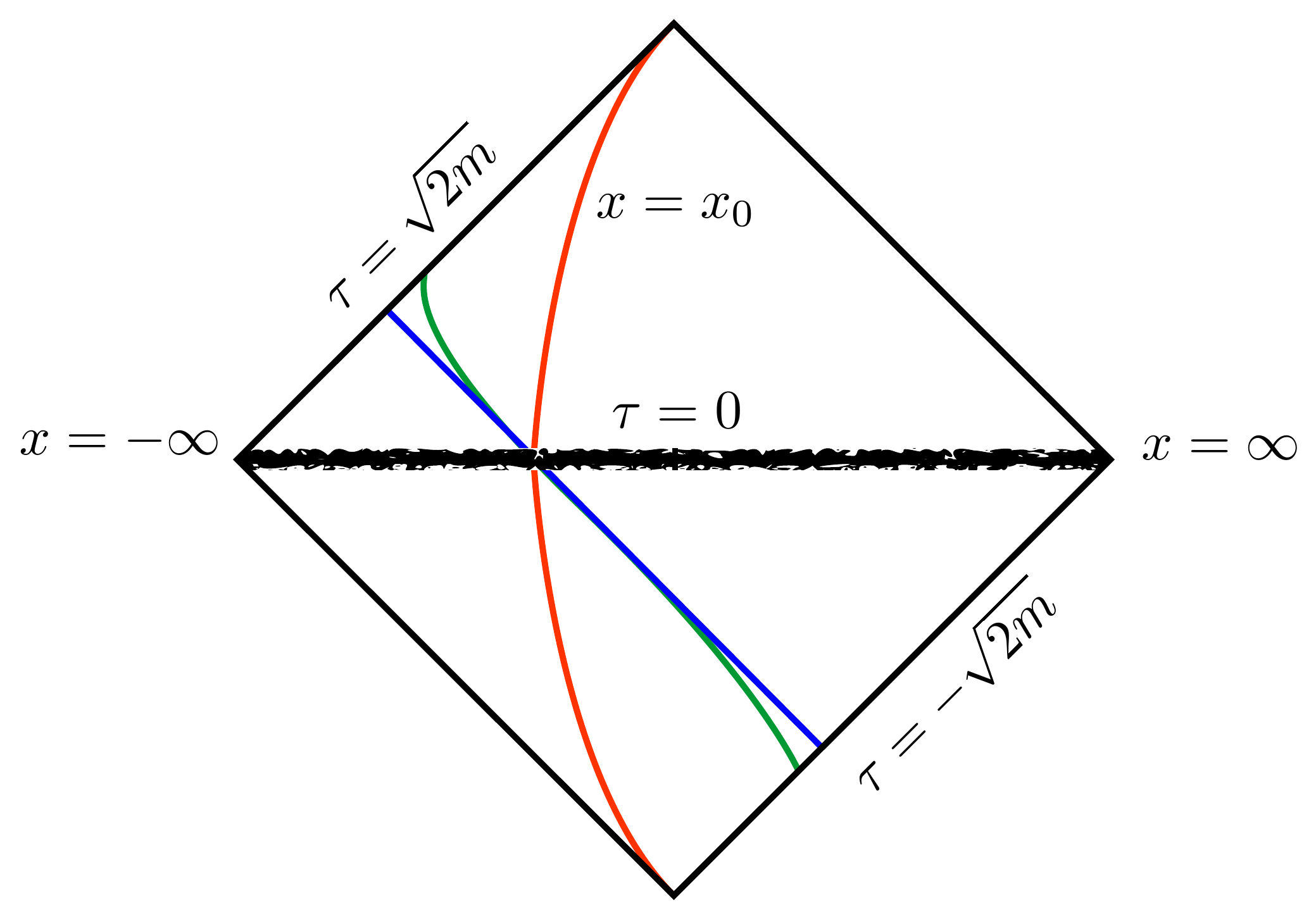}
\caption{\em Illustration of null (blue line) and time-like (green curve with $E<0$) geodesics with $L=0$. The geodesics start in the black hole region (lower part of the diamond), cross the singularity, and continue into the white hole region (top part of the diamond).}
\label{fig:Diamond}
\end{figure}

We first consider the case of a null particle (a photon) falling into the black hole with vanishing angular momentum: $L=0$. The general motions~\eqref{eq:Motions} reduce to the simpler form
\begin{align}\label{eq:Null_Case}
	x(\tau) &= x_0 \pm   \,2 \int\frac{\vert\tau\vert^3}{2m-\tau^2}\,\dd \tau\notag\\
	\phi(\tau) &= \phi_0. 
\end{align}
The  signs derive from the sign of $P$ and correspond to the null geodesics coming from the left or from the right (see also Figure~\ref{fig:ExtSpacetime}).  The integral gives 
\begin{align}\label{eq:NullXP}
	x(\tau) = x_0 \mp   \textsf{s}_\tau\,\left[\tau^2 + 2m\log\left(1-\frac{\tau^2}{2m}\right)\right],
\end{align}
with $\textsf{s}_\tau := \sign{\tau}$ for notational convenience. This solution is regular for all $\tau\in\ ]-\sqrt{2m}, \sqrt{2m}\,[$. These null geodesics start at $x=\mp  \infty$ and end at $x=\pm  \infty$, while intersecting the surface $\tau=0$ at $x = x_0$. See the blue line in Figure~\ref{fig:Diamond}. 

The equations of motion for time-like geodesics with zero angular momentum 
\begin{align}\label{eq:TimeLike_Case}
	x(\tau) &= x_0 { + }  \int\frac{2\, \tau^4 E}{(2m-\tau^2)^\frac{3}{2}\sqrt{1+\frac{\tau^2}{2m-\tau^2}E^2}}\,\dd\tau \notag\\
	\phi(\tau) &= \phi_0,
\end{align}
can also be integrated explicitly yielding the solution
\begin{align}\label{eq:TimeLikeXP}
	x(\tau) =& x_0 { + }   \Bigg[4m\,\artanh{\frac{E \tau}{\sqrt{2m+(E^2-1)\tau^2}}} \notag\\
	&+\frac{2m\,(3-2E^2)E}{(E^2-1)^\frac{3}{2}}\arsinh{\sqrt{\frac{E^2-1}{2m}}\tau} \notag\\
	&-\frac{E\tau\sqrt{(E^2-1)(2m+(E^2-1)\tau^2)}}{(E^2-1)^\frac{3}{2}}\Bigg].
\end{align}
As in the null case, the solution is well-defined in $\tau=0$. What seems to be more worrisome is the parameter range $\vert E \vert  \leq 1$: For $\vert E\vert \rightarrow 1$ the solution~\eqref{eq:TimeLikeXP} seems to be divergent and for $\vert E \vert < 1$ some terms become complex. However, we show in Appendix~\ref{appendix:Limits} that the imaginary terms cancel rendering~\eqref{eq:TimeLikeXP} real also in the parameter range $\vert E \vert < 1$. Moreover, we show that the $\vert E\vert\rightarrow 1$ limit exists and is given by
\begin{align}\label{eq:P_One_Limit}
	x(\tau) = x_0 { \pm }   \left[4m\, \artanh{\frac{\tau}{\sqrt{2m}}-\frac{2\,\tau^3}{3\sqrt{2m}}-2\sqrt{2m}\,\tau} \right].
\end{align}
We also prove in Appendix~\ref{appendix:Limits} that in the $\vert E\vert \rightarrow \infty$ limit the solution~\eqref{eq:TimeLikeXP} converges to the null solution~\eqref{eq:NullXP}
\begin{align}\label{eq:P_Infty_Limit}
	\lim_{\vert E\vert  \rightarrow \infty} x(\tau) = x_0 { \mp }   \textsf{s}_\tau\left[\tau^2+2m\log\left(1-\frac{\tau^2}{2m}\right)\right],
\end{align}
which is exactly what one would expect intuitively. 

\subsection{$E=0$}\label{ssc:PZero}
Under the assumption of vanishing $E$ and arbitrary $L\in\mathbb R \backslash\{0\}$, the equations of motion for null geodesics read
\begin{align}\label{eq:P0Equations}
	x(\tau) &= x_0\notag\\
	\phi(\tau) &= \phi_0 \pm  \,2 
	\int \frac{\dd\tau }{\sqrt{2m-\tau^2}},
\end{align}
where now the sign is determined by the sign of $L$.
The above integral is elementary and yields
\begin{align}
	\phi(\tau) = \phi_0 \pm   2 \, 
	\arctan\frac{\tau}{\sqrt{2m-\tau^2}}.
\end{align}
We see that this solution is as well regular in $\tau=0$. Moreover, the limit $\vert\tau\vert\rightarrow\sqrt{2m}$ exists and is found to be
\begin{align}
	\lim_{\vert \tau \vert \rightarrow \sqrt{2m}} \phi(\tau) = \phi_0 \pm \textsf{s}_\tau \pi.
\end{align}
This means that in the interval $]-\sqrt{2m}, \sqrt{2m}[$ the angular change is $2\pi$. 

Interestingly, the trivial solution $x(\tau) = x_0$ is not as innocuous as it appears at first sight. A generic $x(\tau) = x_0$ curve is not a straight line at $45^\circ$ in a Carter-Penrose diagram. Rather, it looks like the red curve in Figure~\ref{fig:Diamond}, which describes a time-like $E=L=0$ geodesic. The only $x(\tau)=x_0$ lines which are null are obtained by sending $x_0\rightarrow\pm  \infty$. We are therefore led to conclude that $E=0$ null geodesics are confined to the horizons. 

The $E=0$ equations of motion for time-like geodesics turn out to be not integrable in closed analytical form. It is nevertheless possible to integrate them numerically without running into any difficulties.

\subsection{The general Case}\label{ssc:GeneralCase}
For the general case with $L\neq 0$ and $E\neq 0$ it is not possible to write down closed analytic solutions to the equations of motion \eqref{eq:Motions}. But it is still possible to solve the equations numerically and to understand the behavior of geodesics in a neighborhood of $\tau = 0$ by Taylor expanding the integrands of \eqref{eq:Motions}. This expansion results in the approximate solutions
\begin{align}\label{eq:ApproxSol}
	x(\tau) =& x_0 { + }  \frac{E}{\sqrt{2m}L}\left[\frac{\tau^6}{m} + \frac{3 \tau^8}{4 m^2} + \frac{(15 L^2-16m^2\varepsilon)\tau^{10}}{32 L^2 m^3} \right]\notag\\
	&+\mathcal{O}\left(\tau^{11}\right)\notag\\
	\phi(\tau) =& \phi_0 { + }   \frac{1}{\sqrt{2m}}\left[2\tau + \frac{\tau^3}{6m} + \frac{\tau^5}{80}\left(\frac{3}{m^3} - \frac{16 \varepsilon}{L^2}\right)\right. \notag\\
	 &+ \left.\frac{\left(5L^2-16m^2(2E^2+\varepsilon)\right)\tau^7}{448 L^2 m^3}  \right] + \mathcal{O}\left(\tau^8\right).
\end{align}
We observe that both solutions are well-behaved as $\tau\rightarrow 0$ and that there is no problem in crossing the singularity. Moreover, we observe that in both solutions the terms containing an $\varepsilon$, the parameter distinguishing between time-like and null geodesics, is highly suppressed as $\tau\rightarrow 0$. This implies that massive particles approach the behavior of photons the closer they get to the singularity. Notice also the contrast to special relativity: In special relativity, an infinite amount of energy is required to accelerate a massive particle to the speed of light. Hence, only in the limit $\vert E\vert\rightarrow \infty$ does a time-like geodesic approach the behavior of a null geodesic. In the case of a time-like geodesic crossing the Schwarzschild singularity, nothing of the sort is required: Energy is conserved along every time-like geodesic and every $E\neq 0$ time-like geodesic crosses the $r_s=0$ singularity while approaching the behavior of a null geodesic as described by the approximate solution \eqref{eq:ApproxSol}.

\begin{figure}[h]
	\centering
\includegraphics[scale=0.1]{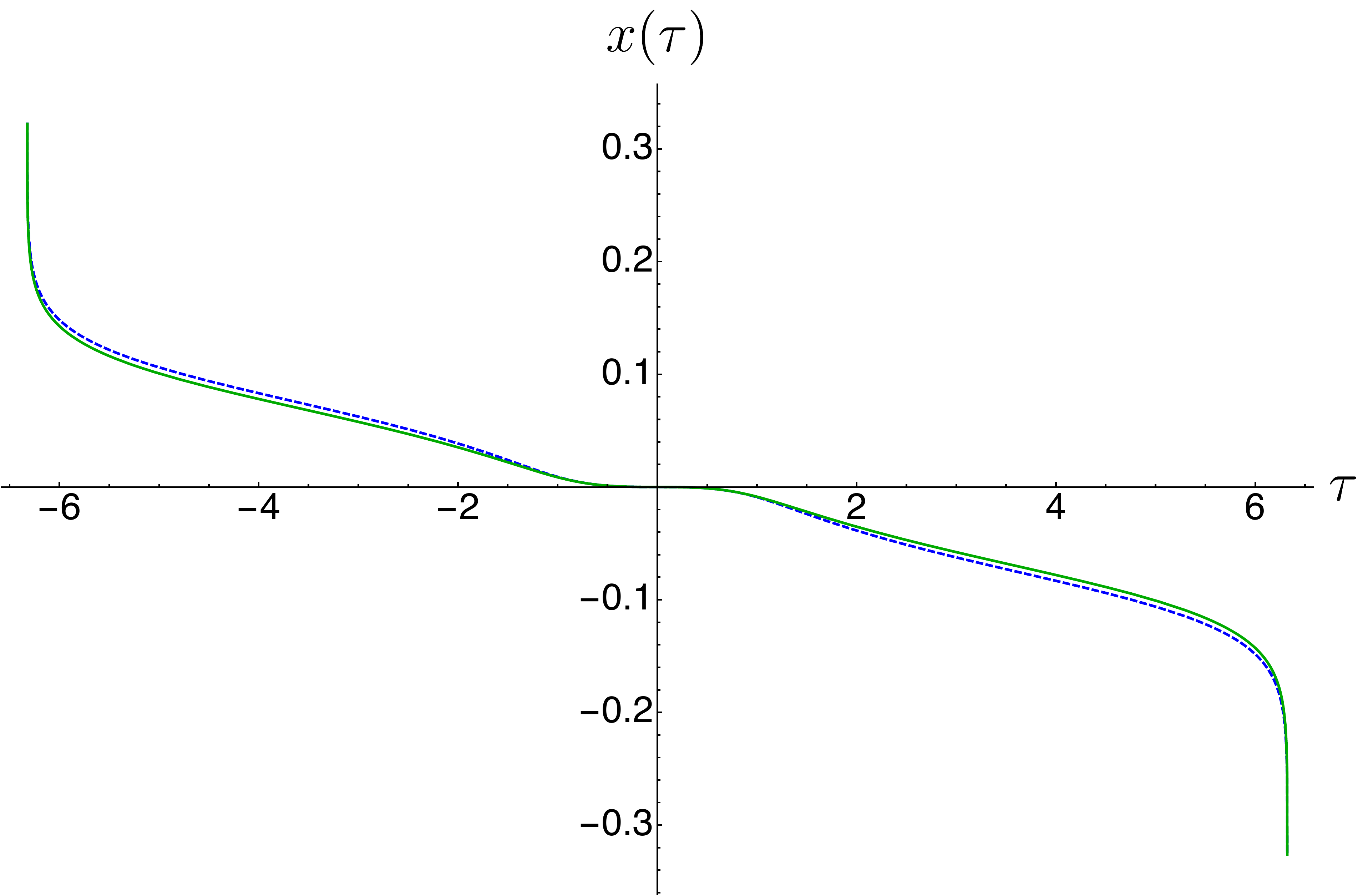}\ \ \ \  \includegraphics[scale=0.1]{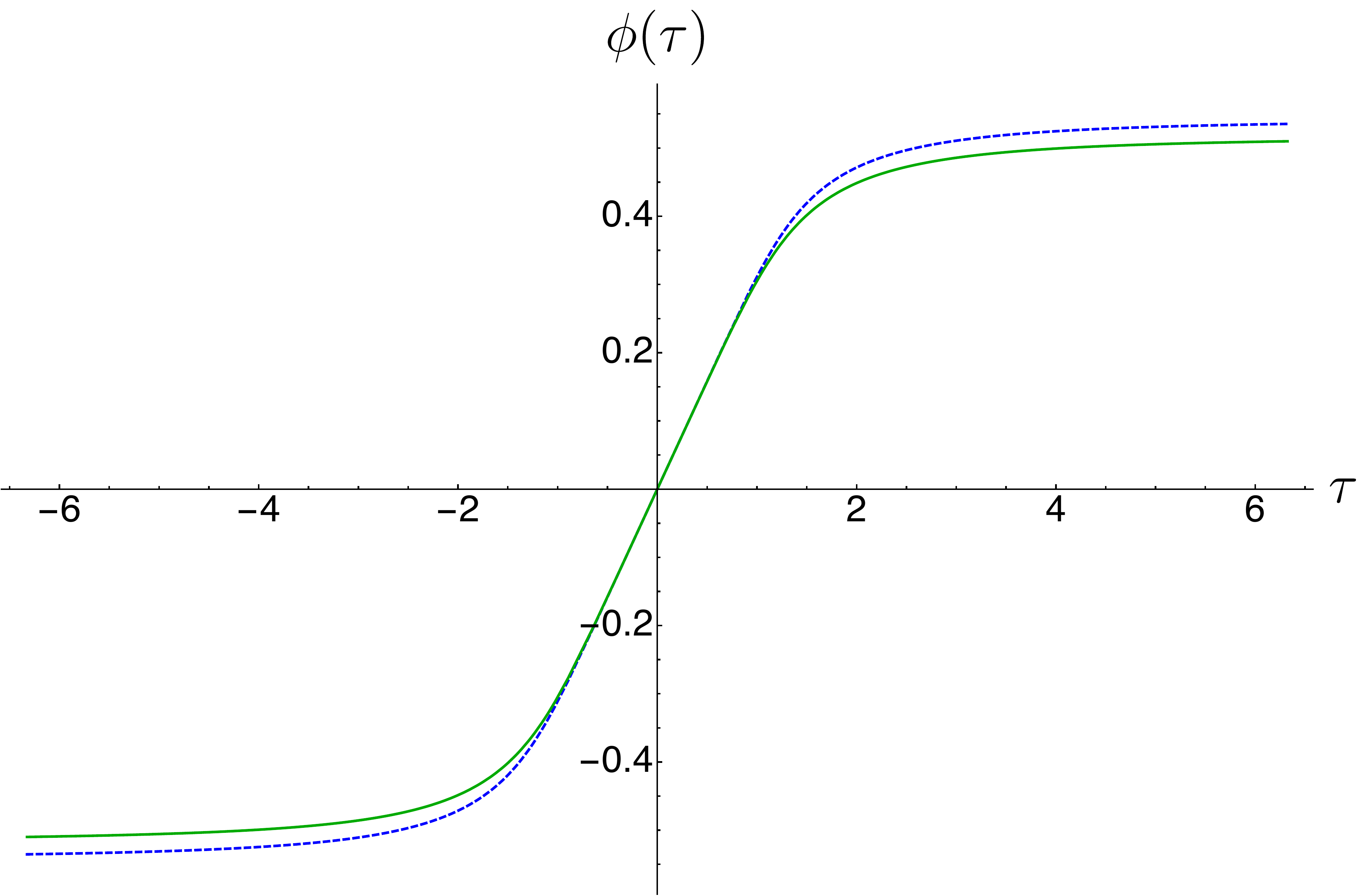}
\caption{\em Integration of~\eqref{eq:Motions} for $x(\tau)$ and $\phi(\tau)$, with $m=20$, $L=-2$, $E=7$. Blue curve: Null solution. Green curve: Time-like solution}
\label{fig:NumericalSolution_Phi}
\end{figure}

For completeness' sake we present a sample solution in Figure~\ref{fig:NumericalSolution_Phi} obtained by numerical integration. 

\section{Quantum gravity around $r_s=0$.}\label{cd} 

The real world is quantum mechanical.  The gravitational field is a quantum field and undergoes quantum fluctuations at small scales. In the real world, therefore, the spacetime metric cannot be everywhere sharp.  A spacetime metric $\dd s_\hbar$ can still be defined in terms of the effective gravitational field, namely the expectation value of $g_{\mu\nu}$ on a quantum state. 

In general, $\dd s_\hbar$ will deviate from the Einstein equation in the vicinity of the classical singularity, because quantum effects are expected to become strong here, and the classical equations of motion are expected to fail; the deviations from an exact solution of the Einstein field equations are parametrized by $\hbar$.  

\begin{figure}[ht]
\includegraphics[height=4.2cm]{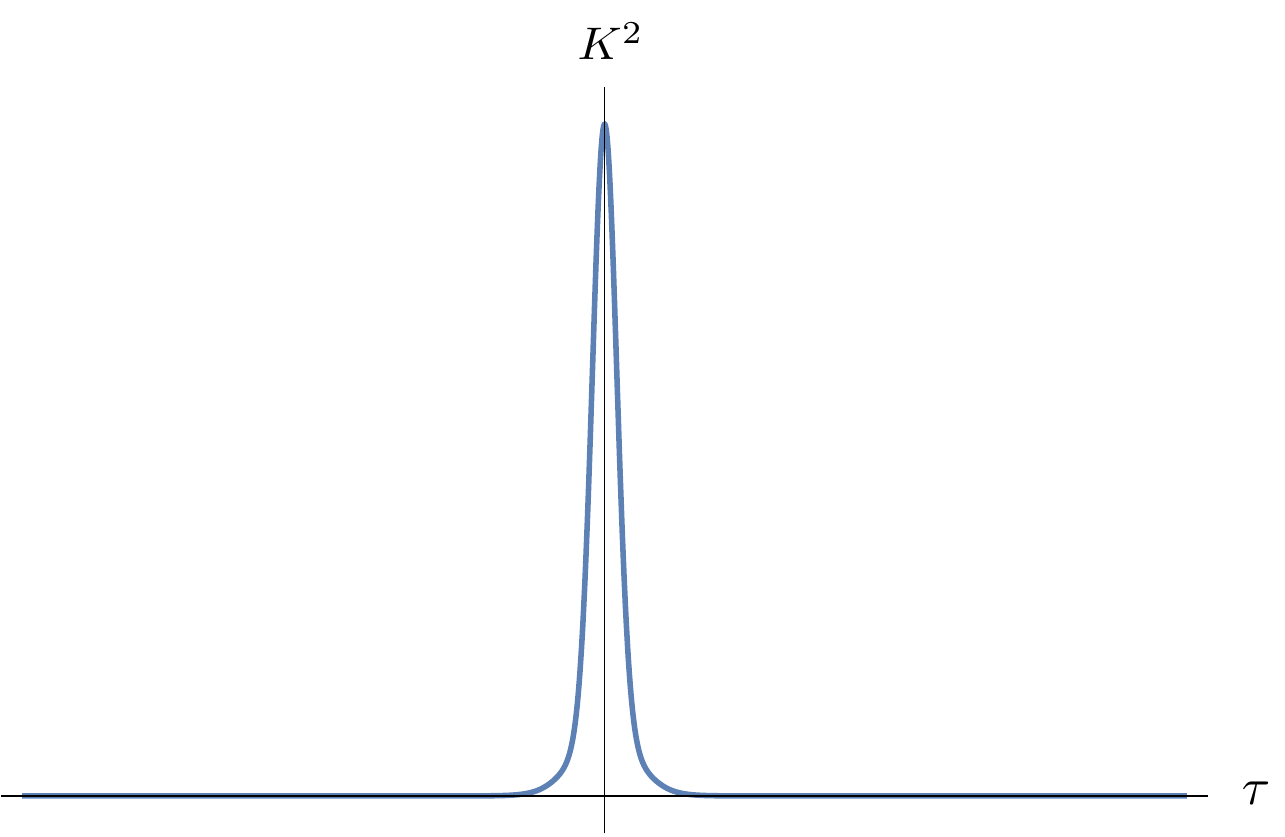}
\vspace{-1em}
\caption{\em The bounded curvature scalar~\eqref{boundedDive}.}
\label{6}
\end{figure}

A simple ansatz for $\dd s_\hbar$ can be obtained replacing $a(\tau)=\tau^2$ in~\eqref{sol} by 
\be
a(\tau)=\tau^2+l_{},
\ee
where $l\!\ll\!m$ is a constant depending on $\hbar$ in a manner that we shall fix soon.  This defines the line element
\be
\dd s_l^2=\frac{4(\tau^2+l_{})^2}{2m-\tau^2}\dd\tau^2-\frac{2m-\tau^2}{\tau^2+l_{}}\dd x^2-(\tau^2+l_{})^2\dd\Omega^2. \label{me2}
\ee 
This line element defines a genuine pseudo-Riemannian space, with no divergences and no singularities. The curvature is bounded (see Figure~\ref{6}). In fact, up to terms of order $\mathcal{O}\left({l_{}}/{m}\right)$ we can easily compute 
\begin{eqnarray}
K^2(\tau)&=& \frac{9\,  l_{}^2- 24\,  l_{} \tau^2+ 48\, \tau^4}{  (l_{} + \tau^2)^8}m^2,
\label{boundedDive}
\end{eqnarray}
which has the \emph{finite} maximum value 
\be
K^2(0)=\frac{9\, m^2}{ l_{}^6} .
\ee
In this geometry the cylindric tube does not reach zero size but bounces at a small finite radius $l_{}$.  The Ricci tensor vanishes up to terms of order~$\mathcal{O}(l_{}/m)$.

The essential point we emphasize in this article is that the $\hbar\to0$ limit of the effective quantum geometry $\dd s_\hbar$ is the geometry~\eqref{met}, depicted in Figure~\ref{4}, and not just its lower half, namely region II of the Kruskal extension.  That is: not a spacetime that ends at a singularity, but rather, a spacetime that crosses the singularity.  The physical relevance of the classical theory is to describe the geometry at scales larger than the Planck scale, and the proper description of the geometry \eqref{me2} at scales much larger than $l$ is {\em a classical spacetime that continues across the central singularity, as described in the first part of this article}.

We can estimate the value of the parameter $l$ from the requirement that the curvature is bound at the Planck scale; we obtain (restoring physical units)
\be
           l\sim l_{Pl}\left({\frac{m}{m_{Pl}}}\right)^{\frac13},
\ee
where $l_{Pl}$ and $m_{Pl}$ is the Planck length and Planck mass. Notice that the bounce away from $r_s=0$ is not at the Planck length, but at a larger scale, defining a ``Planck star" \cite{Rovelli2014}.  

Consider the proper time of a worldline of constant $x$ going all the way from $\tau=-\sqrt{2m}$ to $\tau=+\sqrt{2m}$, crossing $\tau=0$.  Its proper time is 
\be
T=\int_{-\sqrt{2m}}^{\sqrt{2m}} \dd\tau \,\sqrt{\frac{4(\tau^2+l_{})^2}{2m-\tau^2}}=2 \pi \left(m+l_{}\right).  
\ee
In the limit in which $l_{}$ can be disregarded with respect to $m$, a particle following this worldline goes from the Schwarzschild horizon to  $\tau=0$ in a proper time $\pi m$ as predicted by the standard theory, but then continues for another proper time lapse $\pi m$ to the white hole Schwarzschild horizon on the other side of $\tau=0$.  

In the next section, we study an important aspect of the geometry of the effective metric \eqref{me2}.

\subsection{Causal Diamonds crossing $r_s=0$ and their Entropy}\label{cd2} 

The recent article \cite{noi} discusses a solution to the black hole information paradox where quantum gravity effects spark a transition of a black hole into a white hole.  The black hole horizon is then a \emph{trapped} horizon but not an \emph{event} horizon and information that fell into the black hole crosses the transition region and emerges from the white hole.   While the full geometry considered in  \cite{noi} is far more complicated than the geometry considered here, the transition across the $A$ region is the same.

A tentative estimate of the transition probability per unit time for the black-to-white hole tunneling has been computed from covariant loop quantum gravity in  \cite{Christodoulou2018} to be proportional to $\text{e}^{-(m/m_{Pl})^2}$ where $m$ is the mass of the hole at transition time. This makes the transition probable at the end of Hawking evaporation when $m\rightarrow m_{Pl}$. The full evaporation time is $\sim m_o^3$, where $m_o$ is the initial mass of the hole. During the evaporation, the interior volume of the black hole grows, reaching a volume of order $\sim m_o^4$ \cite{Christodoulou2015,Bengtsson2015,Ong2015,Wang2017,Christodoulou2016a}. The quantum transition gives rise to a white hole with small horizon area and large interior volume.  

Remnants in the form of geometries with a small throat and a long tail were called ``cornucopions" in \cite{Banks1992} by Banks \emph{et.al.}~and studied in \cite{Giddings1992c,Banks1993b,Giddings1994,Banks1995}. What was realized in \cite{noi} is that objects of this kind are precisely predicted by conventional classical General Relativity ---white holes with an horizon small enough to be stable--- and are the natural results of the quantum tunneling that ends the life of the black hole.  The large interior volume can  encode a substantial amount of information, despite the smallness of the horizon area. This information is slowly released from the long-lived Planck-mass white hole, purifying the Hawking radiation emitted during the evaporation.

For this scenario to be consistent, the transition region must be large enough to carry the relevant amount of information. In \cite{noi}, an estimate of that amount was given in terms of the interior volume of a preferred foliation.  Here we give a stronger argument, that avoids the non covariance of the choice of the foliation, and is based on Bousso's covariant entropy bound \cite{Bousso1999}.  Bousso's conjecture states that the entropy $S$ on a light-sheet $\mathcal L$ orthogonal to any two-dimensional surface $\mathcal B$ satisfies $S(\mathcal L) \leq A(\mathcal B)/4\hbar$, where $A$ is the area of the surface $\mathcal B$. Here we show that in the crossing region there are closed 2d surfaces with large area satisfying the conditions of Bousso's entropy bound for a large enough entropy to purify the Hawking radiation. 

More precisely, we study the causal diamond defined by two points at opposite sides of the minimal $r_s$ surface: a spacetime point $p = (-\tau_p, x_p, \phi_p, \frac{\pi}{2})$ in the black hole interior (i.e. $0<\tau_p < \sqrt{2m}$) and a spacetime point $p'=(\tau_p, x_p, \phi_p, \frac{\pi}{2})$ in the white hole interior. As~$p'$ lies in $p$'s future, the future light cone of $p$ intersects with the past light cone of $p'$ and hence gives rise to a causal spacetime diamond. In this case, the surface~$\mathcal B$ is given by the intersection of the future and past light cone of $p$ and $p'$ while $\mathcal L$ is the boundary of the causal diamond. 

\begin{figure}[h]
	\centering
\includegraphics[height=4.2cm]{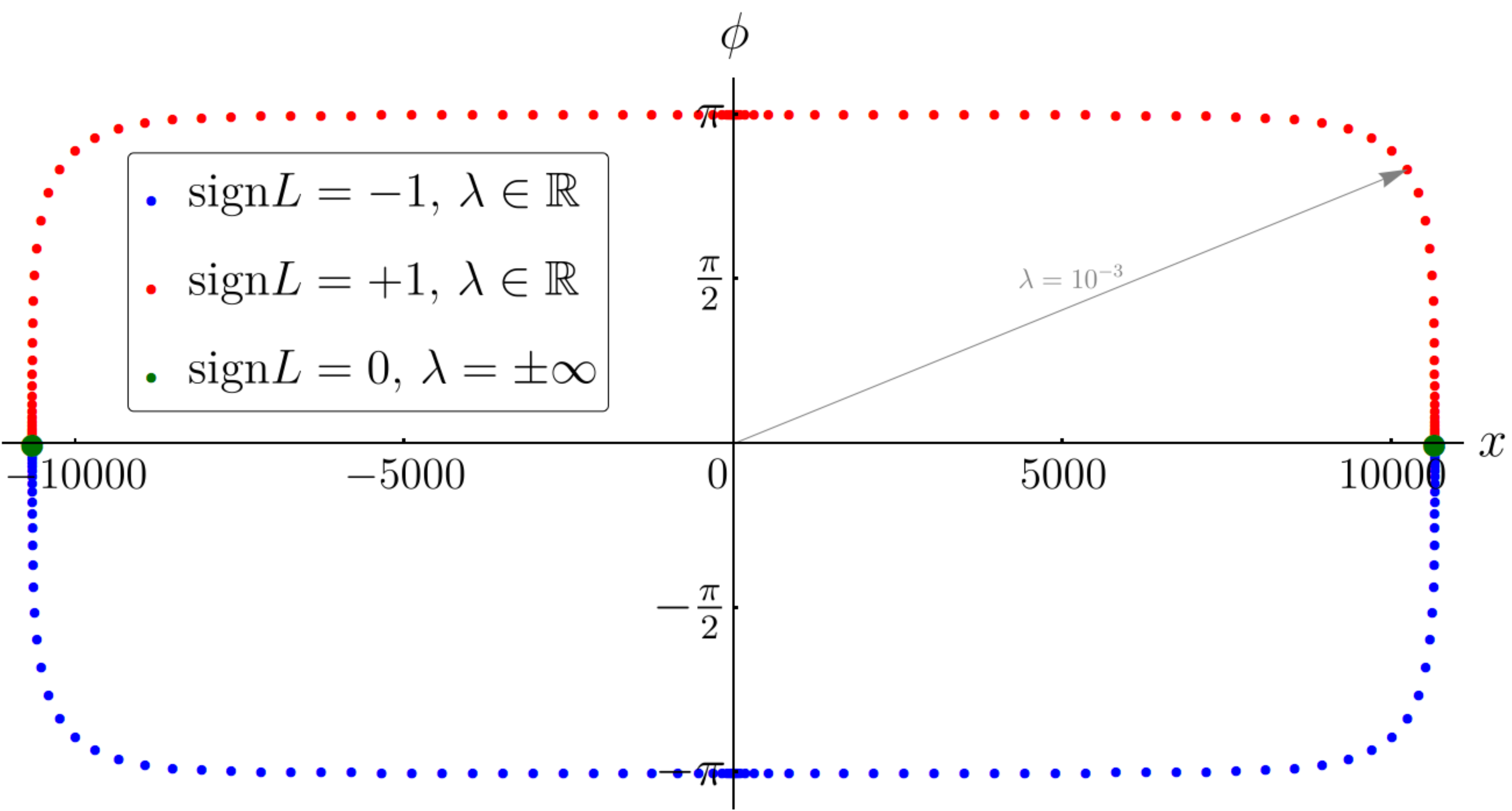}
\caption{Numerical evaluation of the intersection of \eqref{eq:rewritten} with the $\tau=0$ hypersurface for $m=200$ and $\sqrt{2m}-\tau_p = 10^{-11}$. }
\label{fig:ParamCurv}
\end{figure}

The future light cone $\mathcal{I}^+$ of $p$ can be defined as the union of all future null geodesics emerging from that point. Geodesics are labelled by $L$ and $E$ and conservation of angular momentum implies that we can always choose coordinates such that the motion lies in a $\theta = const.$ plane. More precisely, there is always a rotation we can perform to achieve this and therefore it suffices to study in detail the $\theta=\frac{\pi}{2}$ section of $\mathcal{I}^+$ to reconstruct the whole light cone. We can formally write
\begin{align}
	\mathcal{I}^+(p)\big\vert_{\theta=\frac{\pi}{2}} = \bigcup_{L\in\mathbb R}\bigcup_{E\in\mathbb R} (x(\tau), \phi(\tau)), 
\end{align}
where the functions $x(\tau)$ and $\phi(\tau)$ are explicitly given by
\begin{align}\label{eq:orig}
	x(\tau) &= x_p + \int_{-\tau_p}^\tau \frac{2 E\ \tilde{\tau}^6}{(2m-\tilde{\tau}^2)^{\frac{3}{2}}\sqrt{L^2 + \frac{E^2 \tilde{\tau}^6}{2m-\tilde{\tau}^2}}}\dd \tilde{\tau}\notag\\
	\phi(\tau) &= \phi_p + \int_{-\tau_p}^\tau \frac{2 L}{\sqrt{2m-\tilde{\tau}^2}\sqrt{L^2 + \frac{E^2 \tilde{\tau}^6}{2m-\tilde{\tau}^2}}}\dd \tilde{\tau},
\end{align}
where $-\tau_p \leq \tau < \sqrt{2m}$ ensures that the geodesics pass through $p$ and extend into its future. However, different choices of $L$ and $E$ can correspond to the same geodesic and hence there is a lot of redundancy in the above definition of the light cone. To get rid of this redundancy we rewrite $x(\tau)$ and $\phi(\tau)$ as
\begin{align}\label{eq:rewritten}
	x(\tau) &= x_p + \int_{-\tau_p}^\tau \frac{2 \lambda \tilde{\tau}^6}{(2m-\tilde{\tau}^2)^{\frac{3}{2}}\sqrt{1 + \frac{\lambda^2 \tilde{\tau}^6}{2m-\tilde{\tau}^2}}}\dd \tilde{\tau}\notag\\
	\phi(\tau) &= \phi_p + \int_{-\tau_p}^\tau \frac{2\, \sign{L}}{\sqrt{2m-\tilde{\tau}^2}\sqrt{1 + \frac{\lambda^2 \tilde{\tau}^6}{2m-\tilde{\tau}^2}}}\dd \tilde{\tau}.
\end{align}
These equations are obtained from \eqref{eq:orig} by pulling $L$ out of the square root and defining the new parameter $\lambda := \frac{E}{\vert L\vert}$. The advantage is that now it is obvious that all geodesics where $E$ and $L$ have a fixed ratio $\lambda$ and where $L$ has the same sign describe the same geodesic. 
Also, instead of having to build the union over the two continuous parameters $E$ and $L$ to define the light cone we only need to take the union over the continuous parameter $\lambda$ and the discrete values of $\sign{L}$.
\begin{align}
	\mathcal{I}^+(p)\big\vert_{\theta=\frac{\pi}{2}} = \bigcup_{\substack{\lambda \in \mathbb R\\ \sign L = \pm 1}} \bigcup_{\substack{\lambda = \pm \infty \\ \sign L = 0}}(x(\tau), \phi(\tau)).
\end{align}
The past light cone $\mathcal{I}^-$ of $p'$ is defined in an analogous manner, the only difference being the replacement of $-\tau_p$ with $\tau_p$ and the interchange of the integration boundaries in \eqref{eq:rewritten}.
Due to the symmetrical set up, the intersection surface $\mathcal{B}:=\mathcal{I}^+(p)\cap\mathcal{I}^-(p')$ lies on the $\tau=0$ hypersurface and the shape of its cross section is determined by \eqref{eq:rewritten} by setting $\tau=0$ and performing the integrals for all values of $\lambda\in\mathbb R$. This gives two parametric curves in the $x$-$\phi$-plane, one for $\sign L = -1$ and an other one for $\sign L = +1$. They are joined together by the special points $\lambda = \pm\infty$ with $\sign L =0$. Incidentally, these two points simply correspond to the solution discussed in subsection \ref{ssc:LZero} and are explicitly given by $(\phi_p, x_p\pm\tau_p\pm 2m\log\left(1-\tau_p^2/2m\right))$. There are two other special points we can easily locate in the $x$-$\phi$-plane: $\lambda=0$ with $\sign L=\pm 1$ corresponds to the solution discussed in subsection \ref{ssc:PZero} and we get $(\phi_p\pm 2 \arctan(\tau_p/(2m-\tau^2_p)^\frac{1}{2}),x_p)$. These four special cases determine the ranges over which $\phi$ and $x$ change and as we wish to maximize the surface of intersection, we should maximize these ranges. This is achieved by assuming $\tau_p$ to be close to $\sqrt{2m}$, i.e. $\tau_p = \sqrt{2m}-\epsilon$. The range of $x$ is then given by $[-2m\log(\frac{\sqrt{m}}{\sqrt{2}\epsilon}), 2m\log(\frac{\sqrt{m}}{\sqrt{2}\epsilon})]$ and the range of $\phi$ is to very good approximation $[-\pi, \pi]$.\newline\indent
All the other points on the two curves can be determined by numerically evaluating the integrals \eqref{eq:rewritten} for a large range of $\lambda$'s. Figure \ref{fig:ParamCurv} illustrates the result of such a numerical evaluation.

The intersection of geodesics lying in other $\theta=const.$ planes with the $\tau=0$ hypersurface leads to the same elongated sort of rectangle as depicted in Figure \ref{fig:ParamCurv}. The intersection area can therefore be approximated using the regularized metric \eqref{me2} integrated over $[x_{\textsf{min}}, x_{\textsf{max}}]\times[\theta_{\textsf{min}}, \theta_{\textsf{max}}] = [-2m\log(\frac{\sqrt{m}}{\sqrt{2}\epsilon}),2m\log(\frac{\sqrt{m}}{\sqrt{2}\epsilon})]\times[0,\pi]$ for both choices of $\sign L=\pm 1$ and neglecting the $\phi$ contribution to the area (which essentially amounts to neglecting the area of two spheres of radius~$l$).
\begin{align}
	A(\mathcal B) &= \int_{\mathcal B}\sqrt{g\vert_{\mathcal B}}\,\dd^2 \sigma \approx 2\int_{x_{\textsf{min}}}^{x_{\textsf{max}}}\dd x\int_{0}^{\pi}\dd\theta \sqrt{g_{xx}g_{\theta\theta}}\notag\\
	&= 8\pi m\sqrt{2ml}\log\left(\frac{\sqrt{m}}{\sqrt{2}\epsilon}\right).
\end{align}

This area can be made bigger and bigger by taking $\tau_p$ closer to the horizon, but it cannot be made arbitrarily big. The reason is that we can only trust our computations as long as quantum gravity effects are negligible, i.e. as long as we are in region $A$ of Figure \ref{uno}. The finite extent $\Delta x = x_{\textsf{max}}-x_{\textsf{min}}$ of region $A$ has been linked to the lifetime $\tau_{bh}\sim m^3\sim \Delta x$ of the black hole \cite{noi} and yields a finite maximal area of
\begin{align}
	A(\mathcal B) \sim 2\pi \sqrt{2ml}\ m^3 \gg 16\pi m^2.
\end{align}
This result is consistent with the argument given in \cite{noi}.

\section{Conclusion}\label{ssc:Conclusion}

Imagine our technology is so advanced that we can build a spaceship surviving Planckian pressure and we decide to enter the recently found 
17 billion solar mass supermassive black hole in the galaxy NGC 1277~\cite{VanDenBosch2012}.  We of course enter the horizon without any particular bump and start descending.   What happens next? 

Current physical knowledge is insufficient to answer this question.  But the question is well posed in principle and should have a correct answer.  One possibility is  that the world ends at $\tau=0$.  But there is another possibility, which may sound more plausible.  Things can traverse the $\tau=0$ surface and find themselves in the metric of an expanding white hole.  The results of this paper makes this possibility more plausible. 

Whether or not this portion of spacetime is going to be connected to the region outside the black hole depends on the physics of the region $B$ of Figure~\ref{uno}, which requires a more specific use of quantum gravity.  This is discussed elsewhere \cite{noi}. 

\section*{Aknowledgement}

We thank Pierre Martin-Dussaud, Tommaso De Lorenzo and Alejandro Perez for many important exchanges.  CR thanks Tom Banks for pointing out the importance of studying causal diamonds inside the black hole. 

\appendix
\section{Various Limiting Cases}\label{appendix:Limits}
Here we show that the solution~\eqref{eq:TimeLikeXP} is real in the parameter range $\vert E \vert <1$, despite the presence of complex terms. Moreover, we show that the limits $\vert E\vert \rightarrow 1$ and $\vert E\vert \rightarrow\infty$ exist and are given by the equations~\eqref{eq:P_One_Limit} and~\eqref{eq:P_Infty_Limit}, respectively.\newline
To verify that the imaginary part of~\eqref{eq:TimeLikeXP} vanishes we observe that the argument of the artanh function is real and well defined for all values of the parameter $E\in\mathbb R$ since $\tau$ is restricted to the interval $I:=\ ]-\sqrt{2m}, \sqrt{2m}\,[$. We therefore do not need to worry about it. \newline
The last term in the bracket of~\eqref{eq:TimeLikeXP} has, under the assumptions $\vert E\vert < 1$ and $\tau\in I$, a purely imaginary nominator and a purely imaginary denominator. It is therefore, as a whole, a real term. The argument of the arsinh function, on the other hand, is purely imaginary. Using the identity
\begin{align}
	\text{arsinh}\, z &= \log\left(z+\sqrt{1+z^2} \right) \quad\forall z\in\mathbb C
\end{align}
with
\begin{align}
	z=i\,\tau\, \sqrt{\frac{1-E^2}{2m}}=: i\,y\quad y\in\mathbb R,
\end{align}
we deduce
\begin{align}
	\text{arsinh}\,z &= \log\left(i\,y + \sqrt{1-y^2}\right)\notag\\
	&= i\,\text{Arg}\left(i\,y + \sqrt{1-y^2}\right).
\end{align}
Since the Arg-function is real and the term in front of the arsinh is purely imaginary, we find that the middle term in the bracket of~\eqref{eq:TimeLikeXP} is real, too. This shows that the solution~\eqref{eq:TimeLikeXP} is real in $\vert E\vert< 1$.\newline\newline
The simplest way to verify the validity of equation~\eqref{eq:P_One_Limit} is to start from~\eqref{eq:TimeLike_Case} and set $\vert E\vert  = 1$. This results in the integral equation
\begin{align}\label{eq:Simple_Solution}
	x(\tau)= x_0 \pm   \sqrt{\frac{2}{m}} \int \frac{\tau^4}{2m-\tau^2}\,\dd \tau,
\end{align}
which indeed yields
\begin{align}
	x(\tau) = x_0 \pm   \left[4m\, \text{artanh}\frac{\tau}{\sqrt{2m}}-\frac{2\tau^3}{3\sqrt{2m}}-2\sqrt{2m}\,\tau\right].
\end{align}
That this is the same as taking the $\vert E\vert \rightarrow 1$ limit of equation~\eqref{eq:TimeLikeXP} follows from the fact that the integrand of~\eqref{eq:TimeLike_Case} converges uniformly to the integrand of~\eqref{eq:Simple_Solution}. That is, define the functions
\begin{align}
	f_n(\tau) &:=\frac{2 \, \tau^4 \left(1-\frac{1}{n}\right)}{(2m-\tau^2)^\frac{3}{2}\sqrt{1+\frac{\tau^2}{2m-\tau^2}\left(1-\frac{1}{n}\right)^2}} \notag\\
	f(\tau) &:= \sqrt{\frac{2}{m}}\frac{\tau^4}{2m-\tau^2}.
\end{align}
Then,
\begin{align}\label{eq:Uniform_Convergence}
	&\sup_{\tau\in I}\vert f_n(\tau)-f(\tau) \vert\overset{n\rightarrow\infty}{\xrightarrow{\hspace{0.7cm}}} 0\notag\\
	\Longleftrightarrow\quad &f_n\longrightarrow f \text{ uniformly on } I.
\end{align}
The $\vert E\vert\rightarrow 1$ limit of solution~\eqref{eq:TimeLikeXP} now follows suit:
\begin{align}
	\lim_{\vert E\vert\rightarrow 1}x(\tau) &= x_0\pm  \lim_{n\rightarrow\infty}\int f_n(\tau)\,\dd\tau\notag\\
	&= x_0\pm  \int f(\tau)\,\dd\tau.
\end{align}
This is precisely the anticipated result. The $\vert E\vert\rightarrow\infty$ limit of equation~\eqref{eq:TimeLikeXP} can be obtained in a similar manner. To this end, we define the functions
\begin{align}
	g_n(\tau) &:= \frac{2\,\tau^4 n}{(2m-\tau^2)^\frac{3}{2}\sqrt{1+\frac{\tau^2}{2m-\tau^2}n^2}}\notag\\
	g(\tau) &:= 2\,\frac{\vert \tau\vert^3}{2m-\tau^2}.
\end{align}
We recognize the second function to be the integrand of~\eqref{eq:Null_Case}, i.e. the integrand of the null equation of motion with $L=0$. Moreover, one verifies easily that $g_n\longrightarrow g$ uniformly on $I$. We can therefore again exchange limit and integration from which we find for the solution~\eqref{eq:TimeLikeXP}
\begin{align}
	\lim_{\vert E\vert\rightarrow\infty}x(\tau) &= x_0\pm   \lim_{n\rightarrow\infty}\int g_n(\tau)\,\dd\tau\notag\\
	&= x_0\pm  \int g(\tau)\,\dd\tau,
\end{align}
which is precisely the result anticipated in~\eqref{eq:P_Infty_Limit}. We conclude that~\eqref{eq:TimeLikeXP} is real valued and well-defined for all parameter values $E\in\mathbb R\backslash\{-1, 1\}$ and that the limits $\vert E\vert\rightarrow 1$ and $\vert E\vert\rightarrow\infty$ exist and are given by the equations~\eqref{eq:P_One_Limit} and~\eqref{eq:P_Infty_Limit}, respectively.

\bibliographystyle{utcaps}
\bibliography{SchwarzschildGeodesicsv2}
\end{document}